\documentclass{article}
\usepackage{spconf,amsmath,graphicx,hyperref}
\usepackage{cite}
\usepackage{amssymb,amsfonts}
\usepackage{algorithmic}
\usepackage{textcomp}
\usepackage{pgfplots}
\usepgfplotslibrary{groupplots}
\usepackage{xcolor}
\usepackage{enumitem}
\usepackage{booktabs}
\usepackage{multirow}
\usepackage{algorithm}
\usepackage{algorithmic}
\usepackage{caption}
\usepackage[utf8]{inputenc}
\usepackage{hyperref}

\newcommand{\mypar}[1]{\noindent{\bf #1}}

\title{Do You Hear What I Mean? Quantifying the Instruction-Perception Gap in Instruction-Guided Expressive Text-To-Speech Systems}

\name{Yi-Cheng Lin$^1$, Huang-Cheng Chou$^2$, Tzu-Chieh Wei$^3$, Kuan-Yu Chen$^1$, Hung-yi Lee$^4$}
\address{$^1$Graduate Institute of Communication Engineering, National Taiwan University \\[0.1em]
$^2$University of Southern California \quad $^3$University of Michigan \\[0.1em]
$^4$NTU Artificial Intelligence Center of Research Excellence (NTU AI-CoRE)
    }
\begin{document}
\ninept
\maketitle
%

\begin{abstract}
Instruction-guided text-to-speech (ITTS) enables users to control speech generation through natural language prompts, offering a more intuitive interface than traditional TTS. 
However, the alignment between user style instructions and listener perception remains largely unexplored. 
This work first presents a perceptual analysis of ITTS controllability across two expressive dimensions (adverbs of degree and graded emotion intensity) and collects human ratings on speaker age and word-level emphasis attributes.
To comprehensively reveal the instruction-perception gap, we provide a data collection with large-scale human evaluations, named \textbf{E}xpressive \textbf{VO}ice \textbf{C}ontrol (E-VOC) corpus.
Furthermore, we reveal that (1) gpt-4o-mini-tts is the most reliable ITTS model with great alignment between instruction and generated utterances across acoustic dimensions.
(2) The 5 analyzed ITTS systems tend to generate Adult voices even when the instructions ask to use child or elderly voices.
(3) Fine-grained control remains a major challenge, indicating that most ITTS systems have substantial room for improvement in interpreting slightly different attribute instructions.
\end{abstract}
%
\section{Introduction}
\vspace{-5pt}
\label{sec:intro}
Instruction-guided text-to-speech (ITTS) \cite{instructtts, cosyvoice3} enables users to steer speech synthesis using natural-language prompts (e.g., “read this joyfully” or “speak like a child”). This approach offers a transparent and flexible alternative to conventional TTS pipelines \cite{fastspeech2, yourtts, shi2025emotion, breezyvoice} that often require low-level acoustic controls or specialized labels for prosody and timing. By shifting control to free-form language, ITTS promises to enhance accessibility for content creation, assistive technologies, education, and interactive media.

Reliable evaluation is essential for deploying ITTS systems in practical applications. 
Traditional metrics like Mean Opinion Score (MOS) \cite{lo2019mosnet, mittag2021nisqa, highratemos} assess naturalness and Word Error Rate (WER) \cite{arai19_interspeech} measure intelligibility, but these metrics fall short in measuring instructional fidelity, the precise alignment of synthesized speech with fine-grained user prompts. 
This gap raises a central question: \textbf{\textit{Do natural-language instructions for ITTS systems reliably align with listener perceptions, particularly for slightly different attributes like graded emotion intensity?}}

To address this question, we introduce a novel evaluation of ITTS controllability. 
Our study is the first to incorporate adverbs of degree (e.g., slightly, extremely) and graded emotion intensity (e.g., ecstatic, happy) as explicit evaluation dimensions.
We also present the first large-scale collection of human perceptual ratings for speaker age and word-level emphasis.
To systematically examine the gap between instructions and listener perception across these dimensions, we developed a new analysis framework and compiled the \textbf{E}xpressive \textbf{VO}ice \textbf{C}ontrol (E-VOC) corpus\footnote{\href{https://ag027592.github.io/Do-You-Hear-What-I-Mean/}{\texttt{Project Website}}}, consisting of high-quality judgments from over 165 human raters. 
To ensure the reliability of our findings and enable reproducibility, all data were gathered through a quality-controlled process, and we will publicly release both the corpus and the analysis pipeline.

\vspace{-8pt}
\section{Related works and Background}
\vspace{-2pt}
\subsection{ITTS Systems and Selection}
The field of \emph{Instruction-guided Text-to-Speech} (ITTS) has seen rapid advancement, with many models capable of generating speech from descriptive prompts \cite{instructtts}. 
Although robust systems such as Audiobox \cite{vyas2023audiobox} exist, their closed-source nature limits transparency and reproducibility, which are essential for this study. 
Therefore, to ensure a comprehensive and replicable analysis, we selected five representative models across three distinct categories.
First, to represent the state-of-the-art in open-source research, we included \emph{Parler-TTS} \cite{Lyth_2024} and \emph{PromptTTS++} \cite{Shimizu_2024}. 
These models are publicly available, allowing for the in-depth analysis required for this study. 
Second, to represent the leading edge of commercial ITTS systems, we incorporate \emph{GPT-4o-mini-TTS} \cite{openaitts_2025}.
Its efficient and high-quality API provides an insightful analysis for production-grade expressive synthesis.
Finally, to test the capabilities of non-specialized systems, we included \emph{UniAudio} \cite{Yang_2023}, a unified audio generative model. 
Its inclusion allows us to assess whether a general-purpose audio foundation model can achieve perceptual alignment comparable to others.

\vspace{-2pt}
\subsection{ITTS Evaluation Methods}
Prior work has established evaluation methodologies that measure controllability and instruction alignment to evaluate the performance of ITTS systems. 
These approaches can be broadly grouped into three main categories.

\mypar{(1) Attribute-based objective measures.}  
Several studies evaluate ITTS by classifying acoustic or stylistic attributes of the generated speech and comparing them to the prompt.
For example, PromptTTS and PromptTTS 2~\cite{Shimizu_2024,Lyth_2024} measured accuracy in controlling gender, pitch, speed, and volume. 
Building on this idea, our work also includes objective analyses of pitch, speaking rate, and loudness, as they provide interpretable measures of control precision.

\mypar{(2) Embedding-based similarity measures.}  
Other works adopt embedding models to quantify alignment between prompts and audio outputs. 
For example, AudioBox~\cite{vyas2023audiobox} proposed using Joint-CLAP to correlate audio-text embeddings with human judgments of style relevance, while Emosphere~\cite{emosphere}  utilized emotion2vec \cite{emotion2vec} embeddings to evaluate emotion similarity.

\mypar{(3) Instruction-following perceptual measures.}  
A third line of work directly involves human listeners or automated judges to rate how well synthesized speech matches a textual instruction. This category can be divided into two main approaches:
(i) \textit{Human-Centered Evaluation}: This approach treats human perception as the ground truth. For example, InstructTTS~\cite{Yang_2023} introduced a Relevance MOS (RMOS) to score overall prompt alignment, while VoxInstruct~\cite{zhou2024voxinstruct} used a similar Mean Opinion Score for Instruction (MOS-I). Other studies, like EmoVoice~\cite{emovoice}, have focused more narrowly, collecting listener ratings on specific dimensions such as overall expressiveness.
(ii) \textit{Automated Evaluation}: Recent studies have developed automated methods to increase scalability and reduce cost. SpeechCraft~\cite{Jin_2024}, for instance, fine-tuned classifiers to predict attributes like speaker age and word-level emphasis, and used the classifiers to measure how well these predictions aligned with the original instructions. 
Pushing this further, InstructTTSEval~\cite{Huang_2025} benchmarked ITTS systems using Gemini~\cite{Comanici_2025}, which evaluated alignment across various prompts, from low-level acoustic details to high-level role-play instructions.

While these methods are essential for assessing general alignment, they mostly provide coarse outcomes such as overall relevance scores or discrete category matches (e.g., age or emotion class).
In contrast, our work directly measures perceptual controllability along graded and expressive dimensions, including adverbs of degree, fine levels of emotional intensity.
Also, prior works on age and emphasis evaluation have only applied automated methods. However, classifier predictions are tied to their training data and may inherit dataset biases, making them unreliable indicators of how listeners actually perceive expressive attributes. In contrast, our study conducts large-scale human evaluations, providing direct perceptual evidence.

\vspace{-2mm}
\section{Evaluation Framework}
\vspace{-2mm}
\label{sec:framewrok}
We designed a comprehensive evaluation framework to investigate the instruction-perception gap in ITTS systematically.  
This framework consists of 3 core components: the control dimensions that define the evaluation tasks (Section \ref{ssec:control_dimension}), the evaluation metrics used to quantify alignment (Section \ref{ssec:evaluation_definition}), and the E-VOC corpus of human perceptual data collected to support the analysis (Section \ref{ssec:dataset}).

\begin{table}[!b]
\vspace{-2mm}
\fontsize{7}{9}\selectfont
\centering
\caption{Adjectives used in the prompt to control the style of generated speech by ITTS systems for the Emotion–Intensity Adjective dimension. The higher the number in the \textbf{Level} row, the higher the degree. Intensity is the emotional intensity of the adjective from \cite{mohammad-2018-word}.}
\vspace{-2mm}
\begin{tabular}{@{}c|ccccc@{}}
\toprule
\textbf{Level}    & \textbf{1} & \textbf{2}           & \textbf{3} & \textbf{4} & \textbf{5}                            \\  \toprule
\textbf{Happy}      & Satisfied  & Content              & Happy      & Overjoyed  & Ecstatic                              \\ 
\textbf{Intensity}  & 0.500  & 0.688              & 0.788      & 0.909  & 0.954                              \\  \midrule
\textbf{Sad}        & Gloomy     & Disappointed         & Unhappy    & Sad        & Heartbroken                           \\
\textbf{Intensity}  & 0.578  & 0.636              & 0.750      & 0.864  & 0.969                              \\  \midrule
\textbf{Angry}      & Upset      & Frustrated           & Irritated  & Angry      & Outraged                               \\
\textbf{Intensity}  & 0.439  & 0.636              & 0.706      & 0.824  & 0.964                              \\  \midrule
\textbf{Surprised}  & Intrigued  & Unexpected            & Amazed     & Stunned    & Surprised                             \\
\textbf{Intensity}  & 0.430  & 0.711              & 0.781      & 0.820  & 0.930                              \\  \bottomrule
\end{tabular}
\label{tab:emotion_intensity}
\end{table}

\vspace{-2mm}
\subsection{Control Dimension}
\label{ssec:control_dimension}
\vspace{-2mm}
To evaluate ITTS controllability, we define 4 control dimensions that serve as the tasks in our study. 
These are grouped into two categories: two novel dimensions designed to measure fine-grained expressivity, and two established dimensions that assess fundamental aspects of speech synthesis.

\vspace{-2mm}
\subsubsection{Proposed Control Dimensions}
\vspace{-1mm}
We introduce two dimensions to evaluate fine-grained controllability, because complex expressions like sarcasm require subtle prosodic shifts. Precise scalar control is thus a prerequisite for high-fidelity synthesis.

\mypar{Task I. Adverbs of Degree (Adv. Deg.)} tests whether models follow degree modifiers such as \textit{slightly}, \textit{very}, and \textit{extremely} to adjust prosody (loudness, pitch, speaking rate) and emotion.
This dimension is important because adverbial scaling provides users with a simple way to control the fine-grained prosodic variation, essential for storytelling and emotional expression.

\mypar{Task II. Emotion–Intensity Adjective (Emo-I.A.)} evaluates whether ITTS systems can express different degrees of the same emotion, using adjectives that represent progressively stronger intensities.
For 4 core emotions (\emph{happy}, \emph{sad}, \emph{angry}, and \emph{surprise}), we selected candidate adjectives from the human-annotated NRC Emotion Intensity Lexicon~\cite{mohammad-2018-word}. We filtered the candidates to include words that appear frequently in common use (at least 1,200 times on Wikipedia \cite{wiki_dump_2023}).
These adjectives were then arranged into sequences of increasing intensity (Table~\ref{tab:emotion_intensity}), such as \emph{Satisfied}, \emph{Content},  \emph{Happy}, \emph{Overjoyed}, and \emph{Ecstatic} for the ``happy" category. 
This task evaluates whether models can transform these ordered adjective sequences into corresponding perceptual scales of emotion intensity as judged by human listeners.

\vspace{-2mm}
\subsubsection{Other Control Dimensions}
\vspace{-1mm}
In addition to our proposed dimensions, we include the following established tasks to provide a more holistic evaluation of the capabilities of the ITTS models.

\mypar{Task III. Speaker Age  (Age)} evaluates a model's ability to synthesize a voice that reflects a specific perceived age group.
We define four distinct categories: Child (ages 4-12), Teenager (ages 13-19), Adult (ages 20-64), and Elderly (ages 65+). 
Since vocal cues change systematically throughout a person's life, accurately reproducing them is essential for practical applications, such as entertainment and education.

\mypar{Task IV. Word-level Emphasis (Emphasis)}
assesses the ability to place prosodic prominence on a specific target word within a sentence using cues like pitch excursion and duration.
Precise emphasis control is critical for mirroring natural human speech patterns and allows systems to draw listener attention to important information and preserve the original communicative intent.

\begin{table}[!b]

\fontsize{8}{9}\selectfont
\centering
\caption{\small The table summarizes details of the E-VOC across 4 tasks. \# means the number. Cohen’s kappa is for inter-rater agreement.}
\vspace{-2mm}
\begin{tabular}{@{\hspace{0.0cm}}c|@{\hspace{0.2cm}}r@{\hspace{0.2cm}}r|@{\hspace{0.2cm}}r@{\hspace{0.2cm}}r@{\hspace{0.2cm}}}
\toprule
Task                  & \textbf{Adv. Deg.} & \textbf{Emo–I.A.} & \textbf{Emphasis} & \textbf{Age} \\ \midrule
\# of Utterances      & 2,880              & 3,600             & 1,440             & 720          \\
\# of Ratings         & 17,482             & 29,295            & 10,811            & 3,597        \\
\# of Workers         & 29                 & 59                & 27                & 10           \\
Ratings/Utterance     & 6.1                & 8.1               & 7.5               & 5.0          \\ \midrule
\# of Check Utterances& 39                 & 39                & 64                & 30           \\
Cohen's kappa         & 0.170              & 0.226             & 0.410             & 0.439        \\
Performance           & 0.898              & 0.832             & 0.411             & 0.590        \\ \bottomrule
\end{tabular}
\label{tab:dataset}
\end{table}

\begin{figure*}[!h]
\centering
\pgfplotsset{compat=1.17}

\begin{tikzpicture}
  \begin{groupplot}[
    group style={group size=3 by 1, horizontal sep=1.3cm},
    width=6.0cm,
    height=3.8cm,
    ymajorgrids=true,
    grid style=dashed,
    every axis plot/.append style={thick},
    cycle list name=black white,
    x tick label style={rotate=45, anchor=east, font=\scriptsize},
    ylabel style={font=\small}
  ]

  \nextgroupplot[
    title={Loudness Level},
    ylabel={Integrated Loudness (LUFS)},
    symbolic x coords={
      extremely loud,
      very loud,
      loud,
      slightly loud,
      slightly quiet,
      quiet,
      very quiet,
      extremely quiet
    },
    xtick=data,
    ymin=-32, ymax=-12,
    legend to name=modelsLegend,          
    legend columns=5,
    legend cell align={left},
    legend style={draw=none, /tikz/every even column/.style={column sep=6pt}, font=\small},
  ]

  \addplot+[mark=o, blue]
    coordinates {(extremely loud,-19.22) (very loud,-19.66)
                 (loud,-19.71) (slightly loud,-19.97)
                 (slightly quiet,-22.61) (quiet,-23.40)
                 (very quiet,-25.26) (extremely quiet,-25.35)};
  \addlegendentry{gpt-4o-mini-tts}

  \addplot+[mark=*, orange]
    coordinates {(extremely loud,-18.83) (very loud,-19.10)
                 (loud,-17.41) (slightly loud,-18.32)
                 (slightly quiet,-18.92) (quiet,-19.22)
                 (very quiet,-18.77) (extremely quiet,-18.43)};
  \addlegendentry{Parler-TTS-large-v1}

  \addplot+[mark=triangle*, green]
    coordinates {(extremely loud,-19.48) (very loud,-18.64)
                 (loud,-17.84) (slightly loud,-18.53)
                 (slightly quiet,-20.25) (quiet,-19.55)
                 (very quiet,-19.63) (extremely quiet,-18.46)};
  \addlegendentry{Parler-TTS-mini-v1}

  \addplot+[mark=square*, red]
    coordinates {(extremely loud,-29.12) (very loud,-29.92)
                 (loud,-28.55) (slightly loud,-30.90)
                 (slightly quiet,-29.99) (quiet,-26.99)
                 (very quiet,-30.52) (extremely quiet,-29.33)};
  \addlegendentry{UniAudio}

  \addplot+[mark=diamond*, violet]
    coordinates {(extremely loud,-17.19) (very loud,-17.09)
                 (loud,-17.25) (slightly loud,-18.87)
                 (slightly quiet,-19.07) (quiet,-18.11)
                 (very quiet,-18.64) (extremely quiet,-18.22)};
  \addlegendentry{PromptTTS++}

  \nextgroupplot[
    title={Pitch Level},
    ylabel={Fundamental Frequency (Hz)},
    symbolic x coords={
      extremely high,
      very high,
      high,
      slightly high,
      slightly low,
      low,
      very low,
      extremely low
    },
    xtick=data,
    ymin=90, ymax=250,
  ]
  \addplot+[mark=o, blue]
    coordinates {(extremely high,238.7) (very high,238.0) (high,200.5)
                 (slightly high,193.3) (slightly low,138.9) (low,124.5)
                 (very low,122.8) (extremely low,127.1)};
  \addplot+[mark=*, orange]
    coordinates {(extremely high,186.5) (very high,196.1) (high,203.6)
                 (slightly high,193.0) (slightly low,112.6) (low,120.4)
                 (very low,121.0) (extremely low,116.3)};
  \addplot+[mark=triangle*, green]
    coordinates {(extremely high,210.5) (very high,193.2) (high,213.3)
                 (slightly high,174.6) (slightly low,125.8) (low,128.0)
                 (very low,146.4) (extremely low,140.1)};
  \addplot+[mark=square*, red]
    coordinates {(extremely high,154.2) (very high,142.8) (high,155.1)
                 (slightly high,169.5) (slightly low,160.7) (low,142.1)
                 (very low,129.8) (extremely low,130.0)};
  \addplot+[mark=diamond*, violet]
    coordinates {(extremely high,158.4) (very high,159.3) (high,158.9)
                 (slightly high,162.3) (slightly low,121.8) (low,101.9)
                 (very low,108.4) (extremely low,106.4)};

  \nextgroupplot[
    title={Speaking Rate},
    ylabel={Speaking Rate (words/s)},
    symbolic x coords={
      extremely fast,
      very fast,
      fast,
      slightly fast,
      slightly slow,
      slow,
      very slow,
      extremely slow
    },
    xtick=data,
    ymin=0, ymax=4.4,
  ]
  \addplot+[mark=o, blue]
    coordinates {(extremely fast,3.736) (very fast,3.486) (fast,3.338)
                 (slightly fast,3.176) (slightly slow,2.541) (slow,2.500)
                 (very slow,2.241) (extremely slow,2.146)};
  \addplot+[mark=*, orange]
    coordinates {(extremely fast,3.533) (very fast,3.550) (fast,3.486)
                 (slightly fast,3.149) (slightly slow,2.466) (slow,2.482)
                 (very slow,2.577) (extremely slow,2.531)};
  \addplot+[mark=triangle*, green]
    coordinates {(extremely fast,3.559) (very fast,3.674) (fast,3.601)
                 (slightly fast,3.263) (slightly slow,2.509) (slow,2.490)
                 (very slow,2.478) (extremely slow,2.631)};
  \addplot+[mark=square*, red]
    coordinates {(extremely fast,1.862) (very fast,1.862) (fast,2.209)
                 (slightly fast,1.879) (slightly slow,1.830) (slow,1.836)
                 (very slow,1.796) (extremely slow,1.943)};
  \addplot+[mark=diamond*, violet]
    coordinates {(extremely fast,2.551) (very fast,2.616) (fast,2.577)
                 (slightly fast,2.580) (slightly slow,2.562) (slow,2.593)
                 (very slow,2.592) (extremely slow,2.543)};

  \end{groupplot}

  \path (group c1r1.south west) -- (group c3r1.south east)
        node[midway, yshift=3.2cm] {\pgfplotslegendfromname{modelsLegend}};

\end{tikzpicture}

\vspace{-5mm}
\caption{\small Loudness (LUFS), pitch (Hz), and speaking rate (words/s) across ITTS models for \textbf{Task I. Adverbs of Degree}. }
\label{fig:tts_three_in_one}
\vspace{-3mm}
\end{figure*}
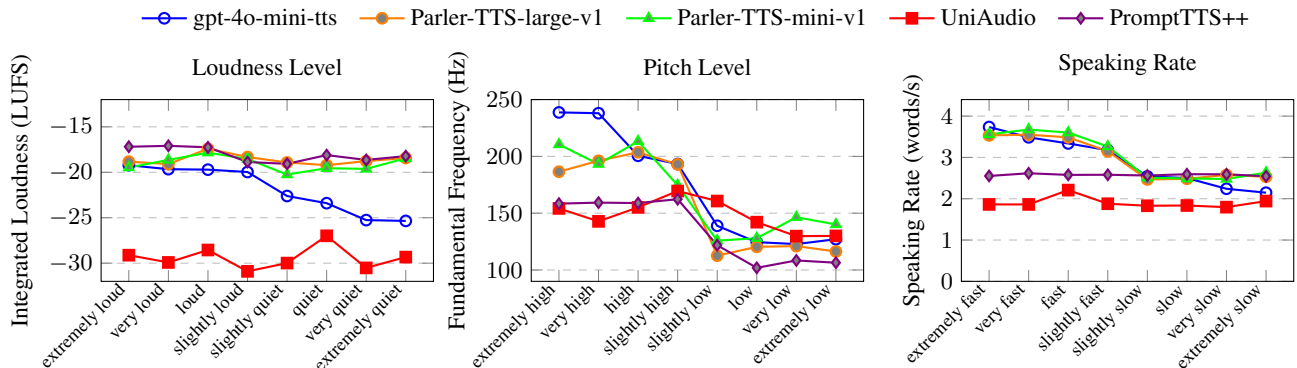

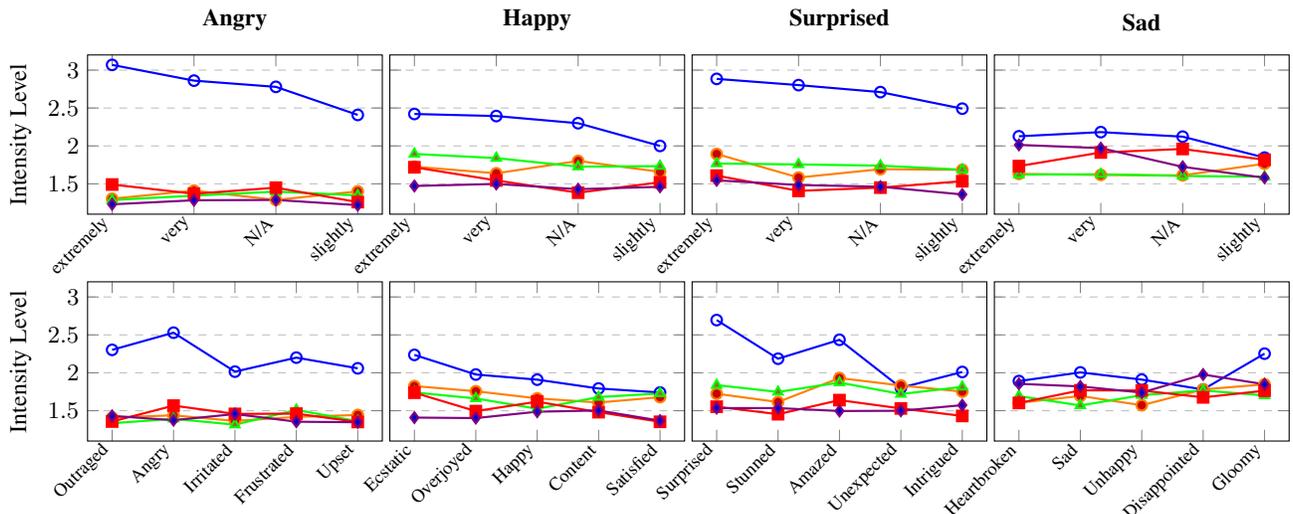
\begin{figure*}[!ht]
\centering
\pgfplotsset{compat=1.17}
\begin{tikzpicture}
  \begin{groupplot}[
    group style={
      group size=4 by 2,
      horizontal sep=0.1cm, 
      vertical sep=0.9cm    
    },
    width=5.5cm,
    height=3.7cm,
    ymajorgrids=true,
    grid style=dashed,
    every axis plot/.append style={thick},
    xtick=data,
    x tick label style={rotate=45, anchor=east, font=\scriptsize},
    ylabel style={font=\small},
    title style={font=\small\bfseries},
    ymin=1.1, ymax=3.2,
    ytick={1.5, 2.0, 2.5, 3.0}, 
  ]


  \nextgroupplot[
    title={Angry},
    ylabel={Intensity Level},
    xticklabels={},
    symbolic x coords={extremely, very, N/A, slightly}, 
    legend to name=modelsLegendEmotions,
    legend columns=5,
    legend cell align={left},
    legend style={draw=none, /tikz/every even column/.style={column sep=6pt}, font=\small},
  ]
  \addplot+[mark=o, blue] coordinates {(extremely,3.0689) (very,2.8612) (N/A,2.7791) (slightly,2.4090)};
  \addlegendentry{gpt-4o-mini-tts}
  \addplot+[mark=*, orange] coordinates {(extremely,1.3038) (very,1.4108) (N/A,1.2880) (slightly,1.4007)};
  \addlegendentry{Parler-TTS-large-v1}
  \addplot+[mark=triangle*, green] coordinates {(extremely,1.2867) (very,1.3444) (N/A,1.3976) (slightly,1.3478)};
  \addlegendentry{Parler-TTS-mini-v1}
  \addplot+[mark=square*, red] coordinates {(extremely,1.4910) (very,1.3690) (N/A,1.4493) (slightly,1.2594)};
  \addlegendentry{UniAudio}
  \addplot+[mark=diamond*, violet] coordinates {(extremely,1.2298) (very,1.2835) (N/A,1.2877) (slightly,1.2188)};
  \addlegendentry{PromptTTS++}

  \nextgroupplot[title={Happy}, yticklabels={}, xticklabels={}, symbolic x coords={extremely, very, N/A, slightly}] 
  \addplot+[mark=o, blue] coordinates {(extremely,2.4214) (very,2.3941) (N/A,2.2996) (slightly,2.0001)};
  \addplot+[mark=*, orange] coordinates {(extremely,1.7268) (very,1.6388) (N/A,1.8021) (slightly,1.6586)};
  \addplot+[mark=triangle*, green] coordinates {(extremely,1.8944) (very,1.8392) (N/A,1.7270) (slightly,1.7336)};
  \addplot+[mark=square*, red] coordinates {(extremely,1.7185) (very,1.5450) (N/A,1.3834) (slightly,1.5203)};
  \addplot+[mark=diamond*, violet] coordinates {(extremely,1.4720) (very,1.4989) (N/A,1.4294) (slightly,1.4611)};

  \nextgroupplot[title={Surprised}, yticklabels={}, xticklabels={}, symbolic x coords={extremely, very, N/A, slightly}] 
  \addplot+[mark=o, blue] coordinates {(extremely,2.8842) (very,2.8023) (N/A,2.7100) (slightly,2.4916)};
  \addplot+[mark=*, orange] coordinates {(extremely,1.8939) (very,1.5829) (N/A,1.6935) (slightly,1.6848)};
  \addplot+[mark=triangle*, green] coordinates {(extremely,1.7680) (very,1.7560) (N/A,1.7398) (slightly,1.6867)};
  \addplot+[mark=square*, red] coordinates {(extremely,1.6086) (very,1.4073) (N/A,1.4500) (slightly,1.5348)};
  \addplot+[mark=diamond*, violet] coordinates {(extremely,1.5462) (very,1.4849) (N/A,1.4619) (slightly,1.3599)};

  \nextgroupplot[title={Sad}, yticklabels={}, xticklabels={}, symbolic x coords={extremely, very, N/A, slightly}] 
  \addplot+[mark=o, blue] coordinates {(extremely,2.1272) (very,2.1822) (N/A,2.1223) (slightly,1.8460)};
  \addplot+[mark=*, orange] coordinates {(extremely,1.6313) (very,1.6174) (N/A,1.6122) (slightly,1.7654)};
  \addplot+[mark=triangle*, green] coordinates {(extremely,1.6221) (very,1.6265) (N/A,1.6038) (slightly,1.5943)};
  \addplot+[mark=square*, red] coordinates {(extremely,1.7340) (very,1.9138) (N/A,1.9601) (slightly,1.8177)};
  \addplot+[mark=diamond*, violet] coordinates {(extremely,2.0136) (very,1.9713) (N/A,1.7244) (slightly,1.5807)};
  

  \nextgroupplot[ylabel={Intensity Level}, symbolic x coords={Outraged, Angry, Irritated, Frustrated, Upset}] 
  \addplot+[mark=o, blue] coordinates {(Upset,2.0591) (Frustrated,2.2000) (Irritated,2.0155) (Angry,2.5291) (Outraged,2.3026)};
  \addplot+[mark=*, orange] coordinates {(Upset,1.4438) (Frustrated,1.4180) (Irritated,1.3650) (Angry,1.4392) (Outraged,1.4101)};
  \addplot+[mark=triangle*, green] coordinates {(Upset,1.3610) (Frustrated,1.5101) (Irritated,1.3157) (Angry,1.3933) (Outraged,1.3347)};
  \addplot+[mark=square*, red] coordinates {(Upset,1.3514) (Frustrated,1.4611) (Irritated,1.4577) (Angry,1.5670) (Outraged,1.3573)};
  \addplot+[mark=diamond*, violet] coordinates {(Upset,1.3474) (Frustrated,1.3548) (Irritated,1.4560) (Angry,1.3747) (Outraged,1.4317)};
  
  \nextgroupplot[yticklabels={}, symbolic x coords={Ecstatic, Overjoyed, Happy, Content, Satisfied}] 
  \addplot+[mark=o, blue] coordinates {(Satisfied,1.7402) (Content,1.7931) (Happy,1.9096) (Overjoyed,1.9773) (Ecstatic,2.2363)};
  \addplot+[mark=*, orange] coordinates {(Satisfied,1.6783) (Content,1.6079) (Happy,1.6627) (Overjoyed,1.7567) (Ecstatic,1.8248)};
  \addplot+[mark=triangle*, green] coordinates {(Satisfied,1.7273) (Content,1.6799) (Happy,1.5253) (Overjoyed,1.6617) (Ecstatic,1.7408)};
  \addplot+[mark=square*, red] coordinates {(Satisfied,1.3535) (Content,1.4807) (Happy,1.6193) (Overjoyed,1.4944) (Ecstatic,1.7374)};
  \addplot+[mark=diamond*, violet] coordinates {(Satisfied,1.3687) (Content,1.5031) (Happy,1.4867) (Overjoyed,1.4028) (Ecstatic,1.4093)};

  \nextgroupplot[yticklabels={}, symbolic x coords={Surprised, Stunned, Amazed, Unexpected, Intrigued}] 
  \addplot+[mark=o, blue] coordinates {(Intrigued,2.0109) (Unexpected,1.8028) (Amazed,2.4352) (Stunned,2.1857) (Surprised,2.6954)};
  \addplot+[mark=*, orange] coordinates {(Intrigued,1.7521) (Unexpected,1.8325) (Amazed,1.9303) (Stunned,1.6144) (Surprised,1.7238)};
  \addplot+[mark=triangle*, green] coordinates {(Intrigued,1.8143) (Unexpected,1.7201) (Amazed,1.8710) (Stunned,1.7478) (Surprised,1.8374)};
  \addplot+[mark=square*, red] coordinates {(Intrigued,1.4309) (Unexpected,1.5298) (Amazed,1.6396) (Stunned,1.4539) (Surprised,1.5527)};
  \addplot+[mark=diamond*, violet] coordinates {(Intrigued,1.5731) (Unexpected,1.4992) (Amazed,1.4960) (Stunned,1.5346) (Surprised,1.5342)};

  \nextgroupplot[yticklabels={}, symbolic x coords={Heartbroken, Sad, Unhappy, Disappointed, Gloomy}] 
  \addplot+[mark=o, blue] coordinates {(Gloomy,2.2523) (Disappointed,1.7801) (Unhappy,1.9104) (Sad,2.0051) (Heartbroken,1.8909)};
  \addplot+[mark=*, orange] coordinates {(Gloomy,1.8478) (Disappointed,1.7802) (Unhappy,1.5721) (Sad,1.6945) (Heartbroken,1.6034)};
  \addplot+[mark=triangle*, green] coordinates {(Gloomy,1.7014) (Disappointed,1.7674) (Unhappy,1.7046) (Sad,1.5690) (Heartbroken,1.6923)};
  \addplot+[mark=square*, red] coordinates {(Gloomy,1.7636) (Disappointed,1.6761) (Unhappy,1.7668) (Sad,1.7687) (Heartbroken,1.6024)};
  \addplot+[mark=diamond*, violet] coordinates {(Gloomy,1.8483) (Disappointed,1.9772) (Unhappy,1.7393) (Sad,1.8202) (Heartbroken,1.8547)};

  \end{groupplot}

\end{tikzpicture}
\vspace{-5mm}
\caption{\small Averaged perceptual emotion intensity of ITTS models across 4 emotions (e.g., \textbf{Angry, Happy, Surprised; Sad}), analyzed by \textbf{Task I. Adverbs of Degree} (top row) and \textbf{Task II. Emotion–Intensity Adjective} (bottom row). The figure shared the same legend with Figure~\ref{fig:tts_three_in_one}.}
\label{fig:tts_emotion_comparison_combined}
\vspace{-4pt}
\end{figure*}

\vspace{-2mm}
\subsection{Evaluation Metrics}
\label{ssec:evaluation_definition}
\vspace{-2mm}
Our framework combines objective acoustic analysis with subjective human perceptual judgments to create a comprehensive evaluation.

\vspace{-2mm}
\subsubsection{Objective Measures}
\vspace{-1mm}
For the \mypar{Adv. Deg.} task, we use objective metrics to quantify changes in the physical properties of the generated speech.
\mypar{Loudness}  is measured in Loudness Units relative to Full Scale (LUFS), following the ITU-R BS.1770-4 standard.
\mypar{Pitch} is calculated as the mean fundamental frequency (F$_0$) per utterance, estimated using the CREPE model \cite{10.1109/ICASSP.2018.8461329}.
\mypar{Speaking rate} is measured in words per second, computed over the utterance duration after trimming leading and trailing silence.

\vspace{-8pt}
\subsubsection{Subjective Measures}
For dimensions requiring stylistic and semantic interpretation, we rely on human perceptual ratings.
\mypar{Emotion Intensity}: For both the Adverbs of Degree and Emotion-Intensity Adjectives tasks, listeners rate the perceived intensity of the target emotion on a 5-point Likert scale.
\mypar{Emphasis}: Listeners use a forced-choice task to identify the most prominent word in an utterance.  
The options include every word in the sentence, plus an \textit{Unclear} option.
\mypar{Age}: Listeners determine the speaker's perceived age by selecting from a forced-choice list: {Child, Teenage, Adult, Elderly, or Unclear\}.

\vspace{-6pt}
\subsection{Human Annotation Collection \& the E-VOC dataset}
\label{ssec:dataset}
To facilitate our human evaluation, we created the E-VOC corpus. 
We generated the audio stimuli for this corpus using five representative ITTS systems: Parler-TTS-large-v1 (\textbf{Parler-large}) \cite{Lyth_2024,Lacombe_2024}, Parler-TTS-mini-v1 (\textbf{Parler-mini}) \cite{Lyth_2024,Lacombe_2024}, PromptTTS++ \cite{Shimizu_2024} (\textbf{Prompt++}), UniAudio \cite{Yang_2023}, and gpt-4o-mini-tts (\textbf{gpt-4o}) \cite{OpenAI_gpt4ominitss}.  

\vspace{-1mm}
\subsubsection{Transcripts Generation}
\vspace{-1mm}
The generation process combined neutral transcripts with specific style prompts. 
We first created eight conversational transcripts for everyday contexts (e.g., teacher-student, customer-server) using Gemini 2.5 Pro~\cite{Comanici_2025}. 
We then paired these transcripts with prompts designed for each control dimension.
For \textbf{acoustic controls}, prompts combined adverbs and adjectives (e.g., ``Speak in a Very High tone").
For the \textbf{Emo-I.A.}, prompts used intensity-specific adjectives (e.g., ``Speak in an Ecstatic tone").
For \textbf{Emphasis}, prompts specified the exact word to be stressed (e.g., ``Articulate clearly, placing special stress on the term 'Sundays'").
For \textbf{Age}, prompts requested a specific age group (e.g., ``Use a Child’s voice").

\subsubsection{Annotation and Quality Control}
We recruited native English speakers from the United States via the Prolific platform. 
All participants completed a brief training session before starting the main annotation task.
To ensure the reliability of our data, we implemented a rigorous quality control protocol. We embedded check utterances with gold-standard labels sourced from public corpora, including CREMA-D \cite{6849440} (for emotion intensity), EMNS (for emphasis), and Nexdata.ai\cite{NEXDATA_2025}/CREMA-D \cite{6849440} (for age). 
We only retained ratings from annotators who demonstrated high accuracy in these check items.
Finally, we report two key reliability metrics in Table~\ref{tab:dataset}: Inter-Rater Agreement, measured using Cohen's Kappa on the check items, and Worker Performance, defined as the percentage of check utterances that each annotator labeled correctly.
At least 5 workers annotate every utterance.

\begin{table}[!t]
\fontsize{7}{9}\selectfont
\centering
\caption{This table summarizes model performance on the \textbf{Age} and \textbf{Emphasis} tasks. Overall performance is reported using accuracy, while class-specific results for the \textbf{Age} task are reported using F1-scores. The best performance in each category is indicated in bold.}
\vspace{-2mm}
\begin{tabular}{@{\hspace{0.2cm}}c@{\hspace{0.2cm}}c@{\hspace{0.2cm}}c@{\hspace{0.2cm}}c@{\hspace{0.2cm}}c@{\hspace{0.2cm}}c@{\hspace{0.2cm}}}
\toprule
\textbf{ITTS}   & \textbf{gpt-4o} & \textbf{Parler-large} & \textbf{Parler-mini} & \textbf{Prompt++} & \textbf{UniAudio} \\ \toprule
\textbf{Task (Metric)}   & \multicolumn{5}{c}{\textbf{Age (Accuracy)}}                                                                          \\ \midrule
\textbf{Overall}  & 0.289           & \textbf{0.294}        & 0.227                & 0.246             & 0.211             \\ \midrule
\textbf{Task (Metric)}   & \multicolumn{5}{c}{\textbf{Age - Class-wise Analysis (F1-score)}}                                                    \\ \midrule
\textbf{Child}    & 0.074           & \textbf{0.113}        & 0.021                & 0.000             & 0.049             \\ 
\textbf{Teenager} & 0.292           & \textbf{0.326}        & 0.149                & 0.127             & 0.148             \\
\textbf{Adult}    & 0.402           & \textbf{0.410}        & 0.337                & 0.330             & 0.281             \\
\textbf{Elderly}  & 0.053           & 0.142                 & 0.199                & 0.310             & \textbf{0.339}    \\ \midrule
\textbf{Task (Metric)}   & \multicolumn{5}{c}{\textbf{Emphasis (Accuracy)}}                                                                     \\ \midrule
\textbf{Overall}  & \textbf{0.265}  & 0.152                 & 0.134                & 0.130             & 0.040             \\ \bottomrule
\end{tabular}
\label{tab:accuracy_3tasks}
\vspace{-4mm}
\end{table}

\begin{table}[!b]
\vspace{-2mm}
\fontsize{8}{9}\selectfont
\centering
\caption{\small Confusion matrices for \textbf{gpt-4o} on \textbf{Age} task. Rows are system outputs (\textbf{Labels}), columns are human judgments. Rows indicate the system-predicted labels, and columns show the categories chosen by human annotators. Higher values on the diagonal represent better alignment between instructions and perception.}
\vspace{-2mm}
\begin{tabular}{lcccc}
\toprule
\multirow{2}{*}{\textbf{Labels}} & \multicolumn{4}{c}{\textbf{Human}} \\
\cmidrule(lr){2-5}
 & Child & Teenager & Adult & Elderly \\
\midrule
Child    & \textbf{7}  & 51  & 121 & 1  \\
Teenager & 0  & \textbf{46}  & 133 & 1  \\
Adult    & 0  & 28  & \textbf{150} & 2  \\
Elderly  & 2  & 10  & 163 & \textbf{5}  \\\bottomrule
\end{tabular}
\label{tab:comnfusion_matrix}
\end{table}

\vspace{-6pt}
\section{Experimental Results and Analyses}
\vspace{-2.8mm}
\subsection{Adverbs of Degree}
\vspace{-1mm}
As shown in Figure \ref{fig:tts_three_in_one}, gpt-4o provides the clearest and most consistent mapping from degree adverbs to acoustic features. Figure~\ref{fig:tts_emotion_comparison_combined} (top row) extends this analysis to perceived emotion intensity under adverb cues.

\mypar{Loudness.} gpt-4o spans a wide LUFS range with a predictable ordering from \textit{slightly} to \textit{extremely.} 
In contrast, both Parler models show limited variation; PromptTTS++ is nearly flat, and UniAudio remains significantly calmer overall.

\mypar{Pitch.} gpt-4o effectively separates \textit{high} and \textit{low} instructions into distinct F$_0$ bands. Other systems exhibit smaller, irregular separations, with degree steps that often compress or overlap.

\mypar{Speaking Rate.} gpt-4o again covers the broadest range with a logical progression from \textit{extremely slow} to \textit{extremely fast,} while other models show minimal or inconsistent changes.


\mypar{Emotion.} Once again, gpt-4o demonstrates strong, consistent gradation within each emotion, with listeners rating \textit{extremely} and \textit{very} prompts as more intense than \textit{slightly.}
The other systems show weaker separation and even occasional reversals.

Overall, gpt-4o is the only model that reliably translates degree modifiers into both the intended acoustic shifts and the corresponding perceptual changes based on our experimental settings.

\vspace{-3mm}
\subsection{Emotion–Intensity Adjectives}
\vspace{-1mm}
gpt-4o was the only system to demonstrate reliable control over graded emotional intensity across all four emotion categories (Figure \ref{fig:tts_emotion_comparison_combined}, bottom row) in this task. 
For instance, listener ratings for gpt-4o increased smoothly along the \textit{Happy} graded emotion words (from \textit{Satisfied} to \textit{Ecstatic}) and the \textit{Surprised} ones (from \textit{Intrigued} to \textit{Surprised}).

Other models show weaker distinctions.
Parler-large and Parler-mini captured some variation, but the perceptual steps between adjacent adjectives were small. 
PromptTTS++ often produced nearly indistinguishable outputs across terms, while UniAudio occasionally exhibited reversed trends, with listeners rating mid-level adjectives as more intense than stronger ones. 
Overall, most ITTS systems can generate distinct categorical emotions, but only gpt-4o reliably controls fine-grained emotion intensity.


\vspace{-1mm}
\subsection{Speaker Age}
\vspace{-1mm}
The \textbf{Age} control task was challenging for all systems, with low overall accuracies reported in Table~\ref{tab:accuracy_3tasks}. 
Parler-large and gpt-4o achieved the highest scores, but only achieved accuracies of 0.294 and 0.289, respectively.
Class-wise F1-score in Table~\ref{tab:accuracy_3tasks} indicates that all models reproduce \emph{adult} or \emph{teenager} speech most reliably, while \emph{child} and \emph{elderly} voices are much harder to generate. 
In particular, gpt-4o achieves the strongest recognition for \emph{teenage} and \emph{adult} prompts, whereas UniAudio and Prompt++ perform relatively better for \emph{elderly} prompts. 
The generation of a child voice was particularly challenging, with extremely low F1-scores across all systems.

Analysis of the gpt-4o confusion matrix (Table~\ref{tab:comnfusion_matrix}) further reveals a strong bias:
regardless of the prompt, listeners most often perceived the output as adult.
This finding suggests that current ITTS models gravitate toward a default adult-like voice with limited control over other age categories.

\vspace{-1mm}
\subsection{Word-level Emphasis}
\vspace{-1mm}
Controlling word-level emphasis was a significant challenge for five ITTS systems. 
As shown in Table~\ref{tab:accuracy_3tasks}, gpt-4o achieved the highest accuracy, yet its score of 0.265 indicates that even the best-performing model struggled. 
Since each sentence contained 6 candidate words on average, random guessing would yield an expected accuracy of only $1/7 \approx 0.143$, including the option “no word emphasized”. 
This task highlights a critical area for improvement, as effective emphasis requires precise and consistent coordination of pitch excursion, duration, and intensity at the word level.


\vspace{-1mm}
\section{Conclusion and Future Work}
\vspace{-1mm}
\mypar{Conclusion.}
This work addresses the largely unexplored link between natural-language instructions and listener perception in ITTS. 
We proposed a novel framework for evaluating fine-grained control using adverbs of degree (e.g., slightly happy) and ordered emotional adjectives (e.g., from \textit{Content} to \textit{Happy} to \textit{Ecstatic}). 
Analysis of five leading ITTS models revealed two key patterns:
(1) Among the systems tested, gpt-4o was the only one that reliably translated degree modifiers and graded adjectives into perceptually ordered changes in loudness, pitch, speaking rate, and emotion intensity.
(2) Fine-grained controls like word-level emphasis and speaker age were inconsistently realized across all models. 
Most ITTS systems defaulted to adult-like voices and produced weak emphasis cues.
To sum up, while current ITTS can follow some high-level styles with coarse reliability, achieving consistent and fine-grained alignment with human perception remains an open challenge.

\mypar{Future Work.}
Our large-scale E-VOC corpus, with its 60,000+ human ratings, provides a valuable resource for developing automated evaluation systems. 
A promising future direction is to use this dataset to train and validate spoken language models  like Gemini \cite{Comanici_2025} as reliable, scalable proxies for human perceptual judgments. 
Developing such an automated judge would significantly accelerate ITTS research by enabling faster and more reproducible evaluation cycles~\cite{Huang_2025,Chiang_2025}.


\section{Acknowledgements}
This work was supported by the Ministry of Education (MOE) of Taiwan under the project Taiwan Centers of Excellence in Artificial Intelligence, through the NTU Artificial Intelligence Center of Research Excellence (NTU AI-CoRE). It was also supported in part by the National Science and Technology Council (NSTC), Taiwan, under Grant No. 114-2917-I-564-030 (to H.-C. Chou). The successful completion of this research was made possible by the academic resources and advanced research infrastructure provided by the National Center for High-Performance Computing, National Institutes of Applied Research (NIAR), Taiwan, and we gratefully acknowledge their invaluable support. We also thank Jie-Wei Jiang for insightful discussions and feedback.

\bibliographystyle{IEEEbib}
\bibliography{refs}
\end{document}